\documentclass[]{sn-jnl}
\usepackage[big]{layaureo}

\usepackage[]{amsmath}
\usepackage[]{amssymb}
\usepackage[]{mathrsfs}
\usepackage{graphicx}
\numberwithin{equation}{section}
\allowdisplaybreaks

\newcommand{\0}{\nonumber}

\newcommand{\dd}{\text{d}}
\newcommand{\e}{\text{e}}
\newcommand{\iu}{{\text{i}\mkern1mu}}
\newcommand{\phibg}{\phi}

\def\be#1\ee{\begin{align}#1\end{align}}

\usepackage{xcolor}
\definecolor{pastgreen}{HTML}{669900}
\definecolor{pastblue}{HTML}{336699}
\definecolor{pastred}{HTML}{990000}
\definecolor{linkcol}{HTML}{663333}
\hypersetup{colorlinks=true,linkcolor={pastblue},citecolor={pastblue},urlcolor={pastblue},pdftitle={Mini boson stars in higher dimensions are radially unstable},pdfauthor={Edgardo Franzin}}

\usepackage{orcidlink}

\usepackage[]{cite}
\usepackage[capitalize]{cleveref}

\makeatletter\g@addto@macro\bfseries{\boldmath}\makeatother%

\begin{document}

\title{Mini boson stars in higher dimensions are radially unstable}

\author{Edgardo Franzin\,\orcidlink{0000-0002-5705-5550}}
\affil{Dipartimento di Fisica, ``Sapienza'' Universit\`a di Roma, p.le Aldo Moro~5, 00185, Rome, Italy}
\affil{INFN, Sezione di Roma1, p.le Aldo Moro~2, 00185, Rome, Italy}
\email{edgardo.franzin@uniroma1.it}

\date{\today}

\abstract{Boson stars are self-gravitating solutions made entirely of fundamental massive bosonic fields.
Here we investigate mini boson stars in $D$ non-compact spacetime dimensions and we show that they are dynamically unstable for $D>4$.}



\maketitle

\section{Introduction}
Numerical arguments strongly suggest that complex massive scalar fields in general relativity may form regular self-gravitating stable configurations known as boson stars~\cite{Jetzer:1991jr,Schunck:2003kk,Liebling:2012fv}.

Boson stars can be classified by the scalar self-interacting potential, and for a fixed boson mass $\mu$, they can come under a variety of sizes and compactnesses --- from below the Chandrasekhar limit for fermion stars, up to supermassive; and compactness that can be negligibly small or very close to that of black holes.

Boson stars could hence be astrophysical stellar objects analogous to neutron stars, or black-hole mimickers, i.e., horizonless objects whose exterior geometry can be very close to that of black holes, able to reproduce some of their observational signatures~\cite{Vincent:2015xta,Barausse:2014tra,Cardoso:2016rao}, while leaving room for detectable deviations~\cite{Macedo:2013qea,Macedo:2013jja,Cardoso:2017cfl}.

Remarkably, four-dimensional boson stars have a stable branch~\cite{Lee:1988av}. Here we investigate the stability of these configurations in higher spacetime dimensions and we prove explicitly that they are always dynamically unstable against radial perturbations.
In four spacetime dimensions, for a number of configurations, the kinetic energy of the star can balance its gravitational energy. Instead, in $D$ spacetime dimensions, as the gravitational energy scales as $1/r^{D-3}$, we do not expect equilibrium, unless (possibly) for a limited region of the parameters space, which gets smaller the higher the number of dimensions.

Therefore, the fact that \emph{higher dimensional} mini boson stars are unstable is hardly surprising, and there are several assertions in the literature, solely motivated by binding-energy arguments~\cite{Hartmann:2010pm,Brihaye:2015jja,Brihaye:2016ibz,Blazquez-Salcedo:2019qrz}.\footnote{$D$-dimensional perturbative equations similar to \cref{deltalambda,deltaphi} below were derived in Ref.~\cite{Astefanesei:2003qy}, but solved uniquely in four-dimensional asymptotically anti-de Sitter spacetimes.}
In this note we provide clear evidence --- see also Ref.~\cite{Franzin:2017gac}.

\section{Background solutions}

General relativistic boson stars are self-gravitating solutions of massive complex scalar fields.
The general action in $D$ spacetime dimensions with $D\geq4$ reads (in geometrised $c=G_D=1$ units)
\be\label{actionBS}
S = \int \dd^Dx\,\sqrt{-g}\left(\frac{R}{16\pi} - \left|\nabla\Phi\right|^2 - V\left(\left|\Phi\right|^2\right)\right).
\ee

We consider a time-independent and spherically symmetric line element, in coordinates $x=\{t,r,\theta_1,\ldots,\theta_{D-3},\theta_{D-2} \equiv \varphi\}$ with $0\leq\theta_i\leq\pi$ for $i=\{1,\ldots,D-3\}$ and $0\leq\varphi\leq2\pi$
\be
\dd s^2 = -\e^{\nu(r)}\,\dd{}t^2 + \e^{\lambda(r)}\,\dd{}r^2 + r^2\,\dd\Omega^2_{D-2}\,,
\ee
where
\be
\dd\Omega^2_{D-2} = 
\sum_{i=1}^{D-2} \Pi(i)\,\dd\theta_{i}^2\,,\quad
\Pi(i) = \begin{cases}1 & i=1\\ \prod_{j=1}^{i-1} \sin^2\theta_j &i>1\end{cases}\,.
\ee

We further assume a harmonic ansatz for the complex scalar field (with a convenient numerical factor)
\be\label{Phi_bg}
\Phi (t,r) = \frac{1}{\sqrt{8\pi}}\,\phibg (r)\,\e^{-\iu\omega t}\,,
\ee
whose energy-momentum tensor $T_{ab}$ read
\be
T_{ab} = \frac{1}{2} \Big[\nabla_{\!a}\Phi\,\nabla_{\!b}\Phi^* + \nabla_{\!a}\Phi^*\,\nabla_{\!b}\Phi
-g_{ab}\left(\left|\nabla\Phi\right|^2 + V\left(\left|\Phi\right|^2\right)\right)\Big].
\ee

In the following with focus on mini boson stars, characterised by the potential $V = \mu^2\left|\Phi\right|^2$, being $\mu$ the mass of the scalar field.

Let $E_{ab} \equiv G_{ab} - 8\pi T_{ab}=0$ be the gravitational field equations.
Because of spherical symmetry, the angular components are proportional to each other, i.e.,
$E_{\theta_i\theta_i} = \Pi(i) E_{\theta_1\theta_1}$ for $i=\{2,\ldots,D-2\}$.
Furthermore, $E_{\theta_1\theta_1}$ is a consequence of $E_{tt}$, $E_{rr}$, and the Klein--Gordon equation (and its complex conjugate).
Therefore, the relevant Einstein--Klein--Gordon field equations are
\be
\lambda' &= \frac{r \phibg'^2}{D-2} + \frac{r \e^{\lambda} \left(\mu^2 + \e^{-\nu}\omega^2\right)\phibg^2}{D-2} - \frac{(D-3) \left(\e^{\lambda}-1\right)}{r}\,,\label{eqlambda}\\
\nu' &= \frac{r \phibg'^2}{D-2} - \frac{r \e^{\lambda} \left(\mu^2 - \e^{-\nu}\omega^2 \right)\phibg^2}{D-2}+\frac{(D-3) \left(\e^{\lambda}-1\right)}{r}\,,\label{eqnu}\\
\phibg'' &= \left(\frac{\lambda'-\nu'}{2} - \frac{D-2}{r}\right) \phibg' + \e^{\lambda} \left(\mu^2 - \e^{-\nu}\omega^2\right)\phibg\,.\label{eqphi0}
\ee

The $\text{U}(1)$ invariance of the action \eqref{actionBS} implies the existence of a conserved current $j_a = \frac{\iu}{2} \left(\Phi \nabla_{\!a}\Phi^* - \Phi^* \nabla_{\!a}\Phi\right)$.
The spatial integral of the time component of this current is the boson number (Noether charge)
\be\label{bosonN}
N_D = \int \dd^{D-1}x\,\sqrt{-g}\, j^0 = \frac{\pi^{\frac{D-3}{2}}}{4\Gamma\!\left(\frac{D-1}{2}\right)}\,\omega \int \dd{}r\,r^{D-2} \e^\frac{\lambda-\nu}{2}\phibg^2\,,
\ee
while the total mass of the configuration is
\be\label{massM}
M_D = \lim_{r\to\infty}\frac{(D-2)\pi^\frac{D-3}{2}r^{D-3}}{8\Gamma\!\left(\frac{D-1}{2}\right)}\left(1-\e^{-\lambda}\right).
\ee
%

With the two above quantities it is possible to define the binding energy as $E_\text{bind}=M_D-\mu N_D$.

Although the scalar profile approaches its asymptotic value exponentially fast, the matter distribution is localised in a radius of order $1/\mu$, yet diffused all over the radial direction.
As a consequence, boson stars do not have a hard surface but it is possible to define an effective radius e.g., as the radius within which the 99\% of $M_D$ is contained.

For a given number of spacetime dimensions $D$, background configurations are obtained by integrating numerically \cref{eqlambda,eqnu,eqphi0} along with boundary conditions. We impose regularity at the origin $r=0$\footnote{In practice, to increase the accuracy of the numerical integration, we consider a higher-order expansion near the origin whereof \cref{origin_bg} represents the first terms.}
\begin{subequations}\label{origin_bg}\be
\nu &= \nu_c\,,\quad \nu' = 0\,,\quad \nu'' = \frac{(D-2) \e^{-\nu_c} \omega^2 - \mu^2}{(D-2) (D-1)}\,\phi_c^2\,,\\
\lambda &= 0\,,\quad \lambda' = 0\,,\quad \lambda'' = \frac{\mu^2 + \e^{-\nu_c}\omega^2}{(D-2)(D-1)}\,\phi_c^2\,,\\
\phibg &= \phi_c\,,\quad \phibg' = 0\,,\quad \phibg'' = \frac{\mu^2-\e^{-\nu_c} \omega^2}{2(D-1)}\,\phi_c\,,
\ee\end{subequations}
whereas at infinity we impose the metric to be asymptotically flat and the scalar field to vanish.
\Cref{eqlambda,eqnu,eqphi0} no longer depend on $\mu$ after the transformations $r\to\mu r$ and $\omega\to\omega/\mu$; this means that in our numerical routine we can set $\mu=1$ without loss of generality.
For each scalar field central value $\phi_c$, the background equations form an eigenvalue problem for the frequency $\omega$, which we solve using a standard shooting method.
The value $\nu_c$ is not univocal, but can always be chosen such that $\e^\nu=1$ at infinity, by an appropriate time coordinate transformation.
In general, the boundary conditions are satisfied by a discrete set of frequencies; here we focus on the ground state, corresponding to a nodeless scalar field.

\begin{figure}[!ht]
\centering
\includegraphics[height=0.27\textwidth]{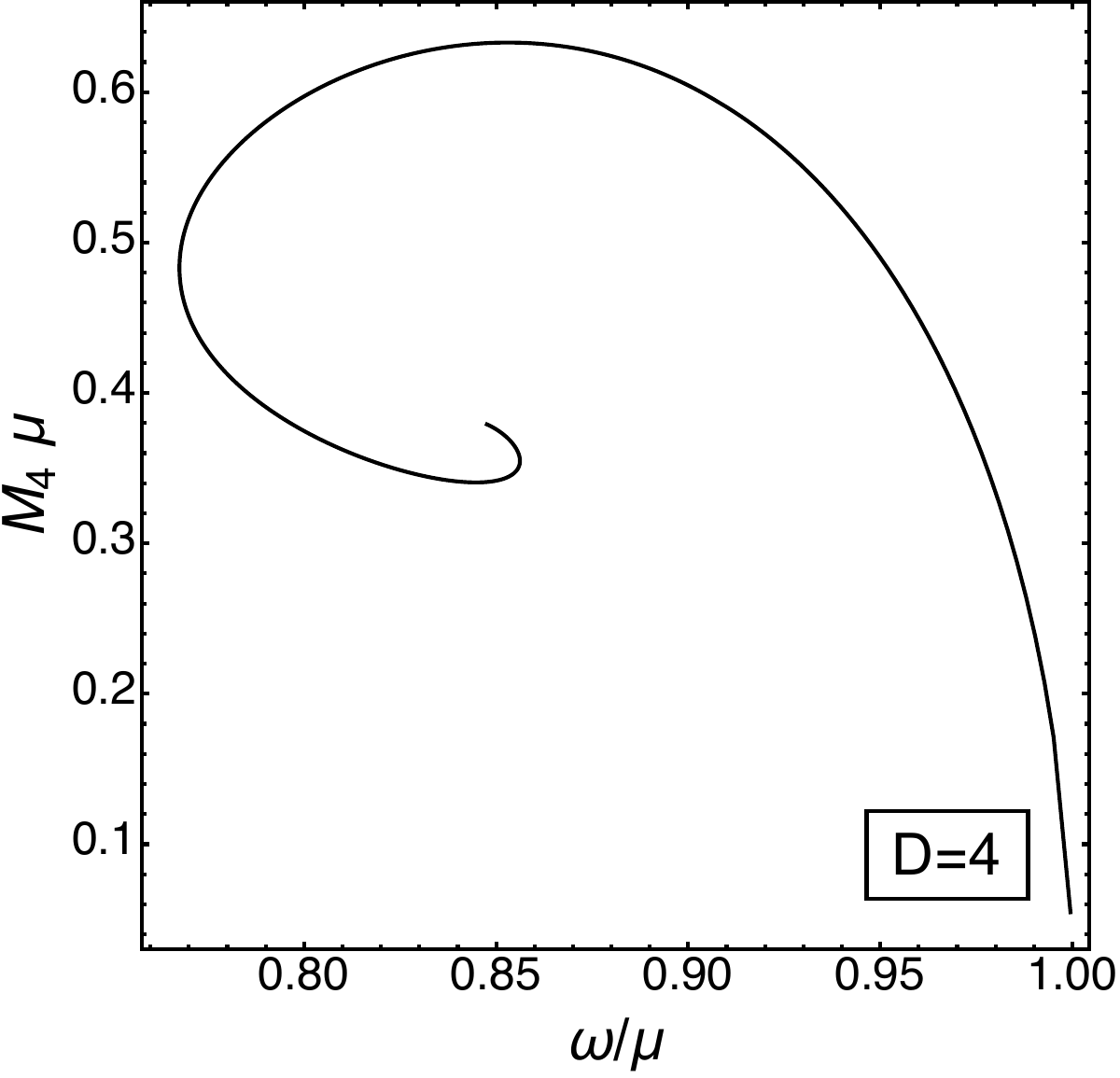}\quad
\includegraphics[height=0.27\textwidth]{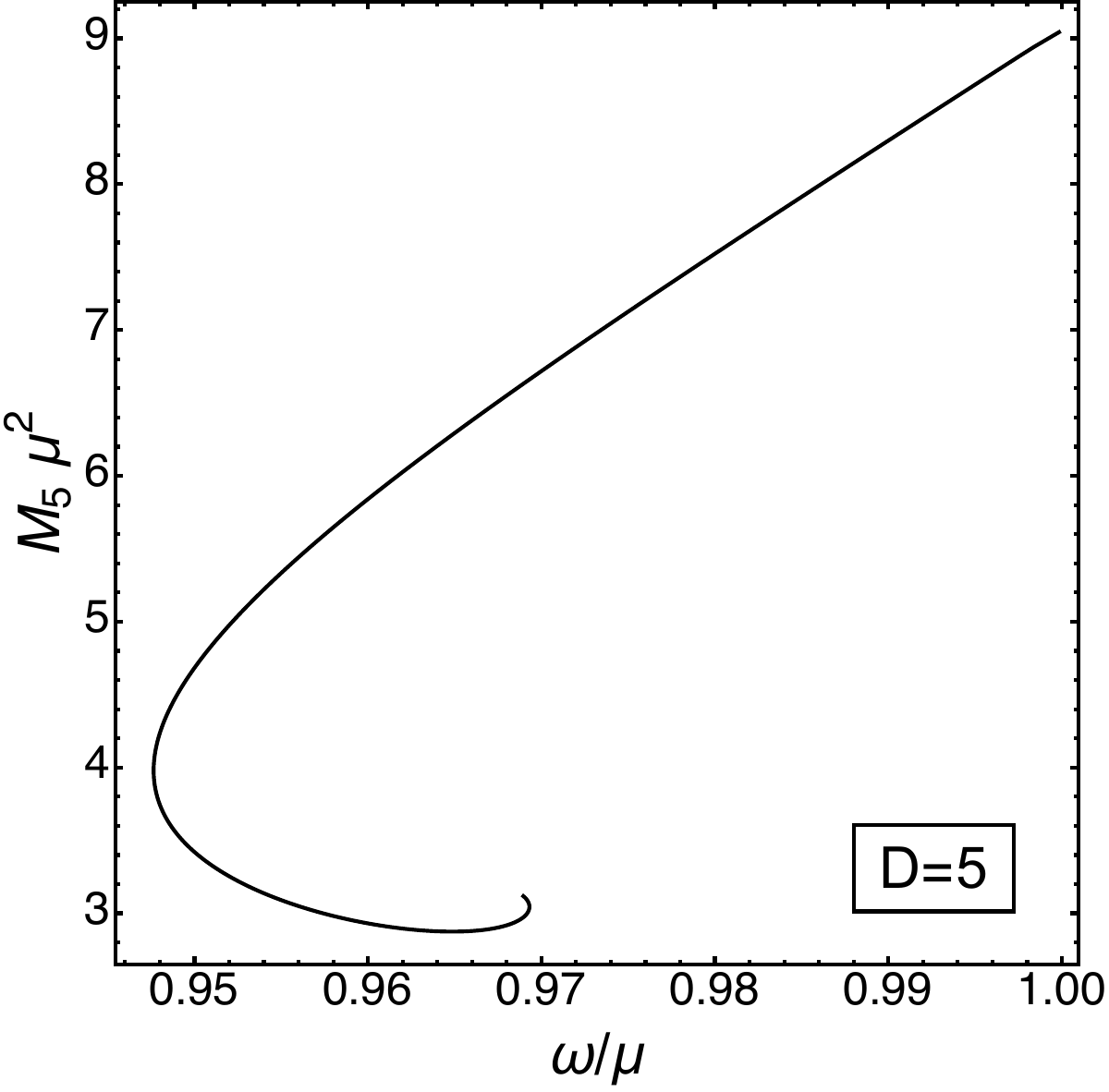}\quad
\includegraphics[height=0.27\textwidth]{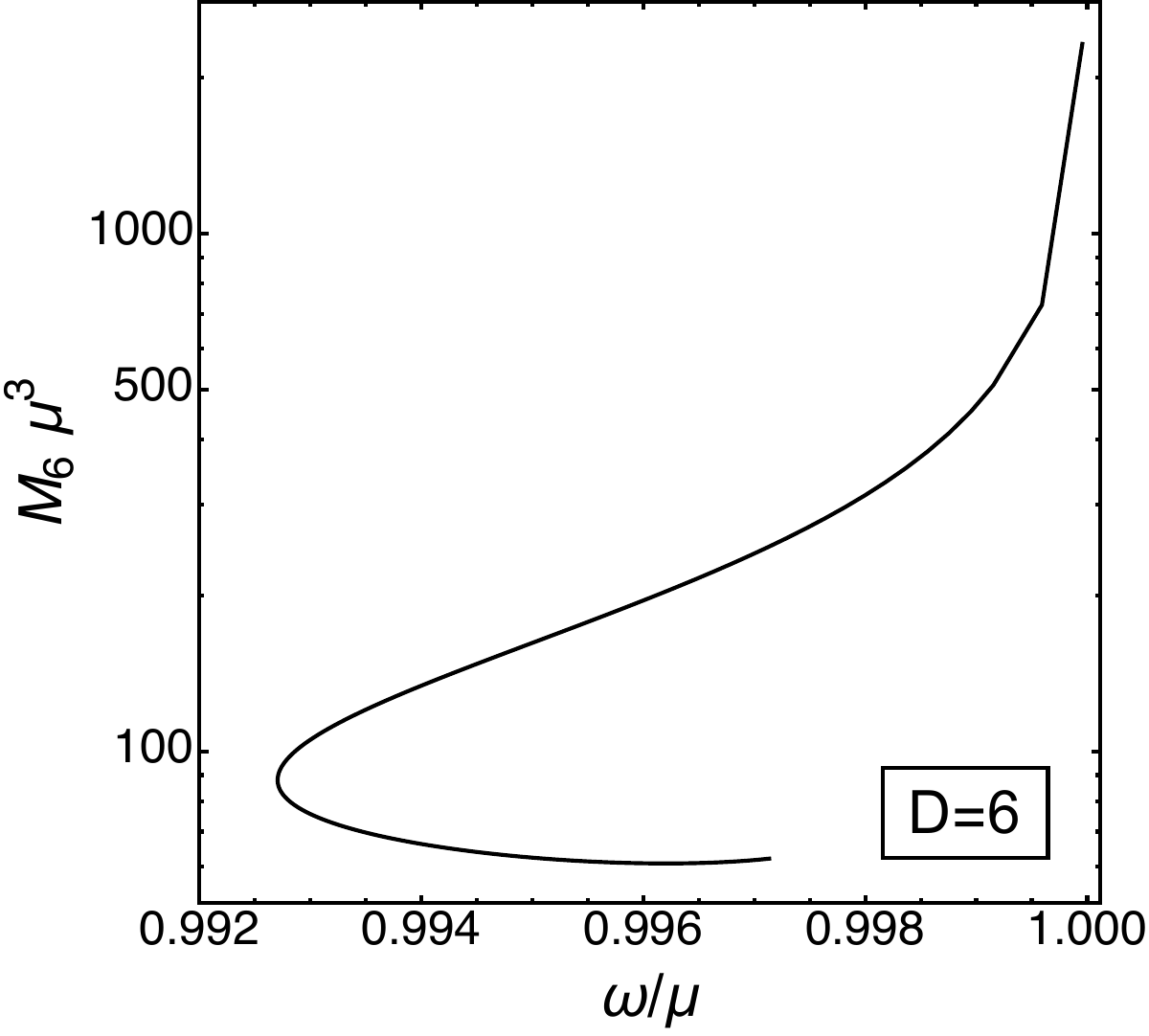}
\caption{Total mass as a function of the frequency of the ground state for mini-boson-star equilibrium solutions in $D=\{4,5,6\}$.}
\label{fig:massfrequency}
\end{figure}

\begin{figure}[!ht]
\centering
\includegraphics[height=0.27\textwidth]{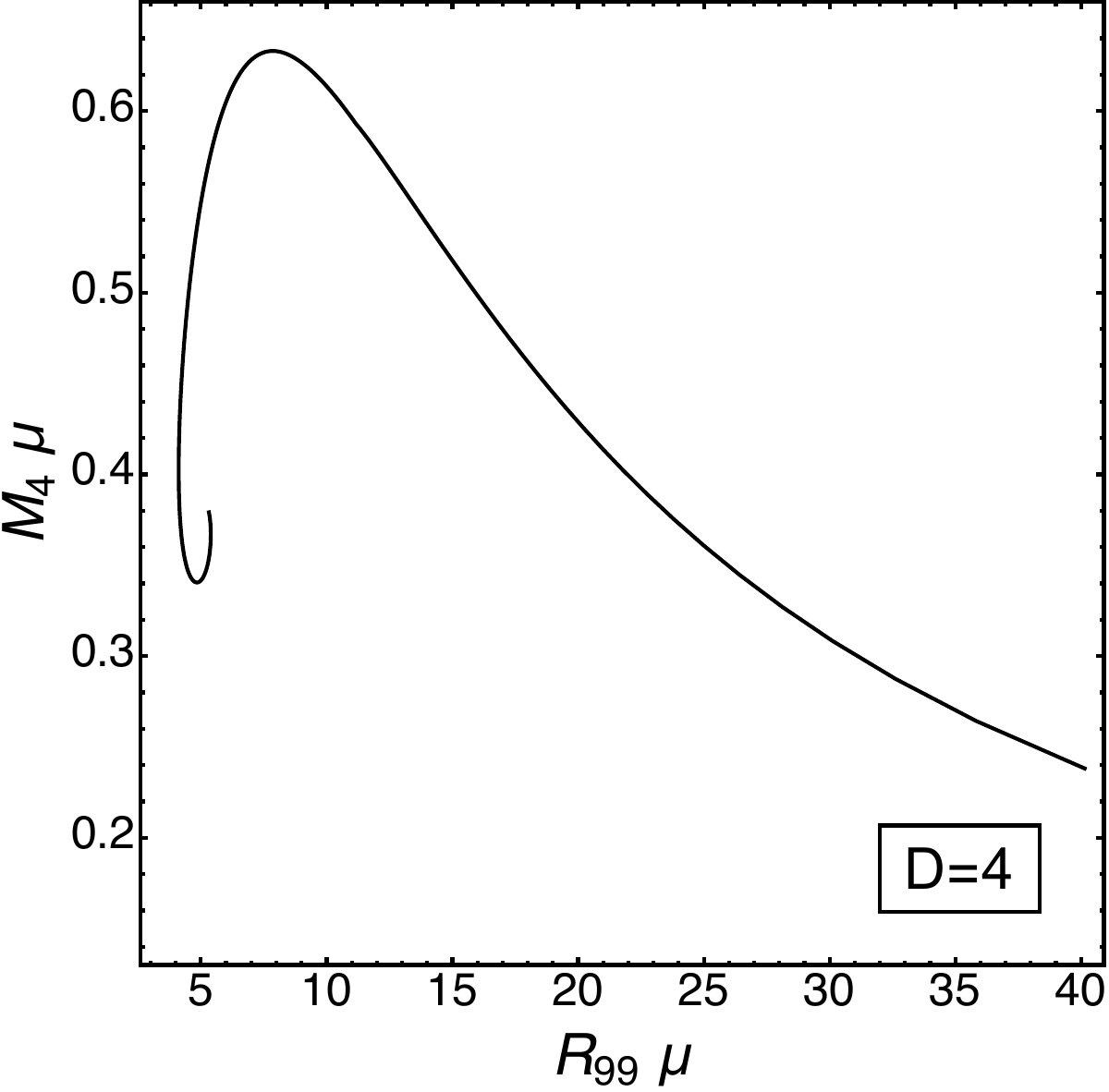}\quad
\includegraphics[height=0.27\textwidth]{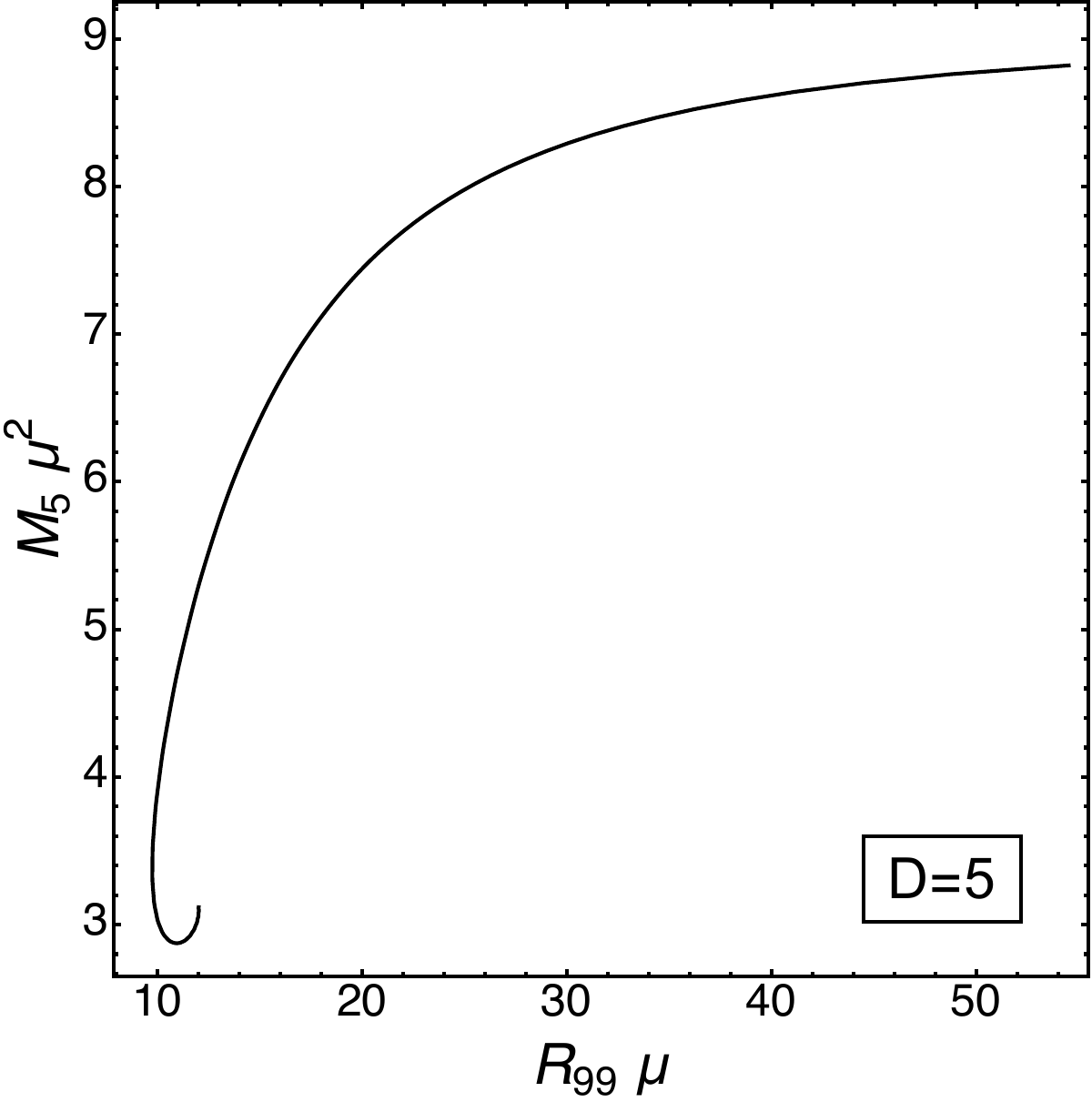}\quad
\includegraphics[height=0.27\textwidth]{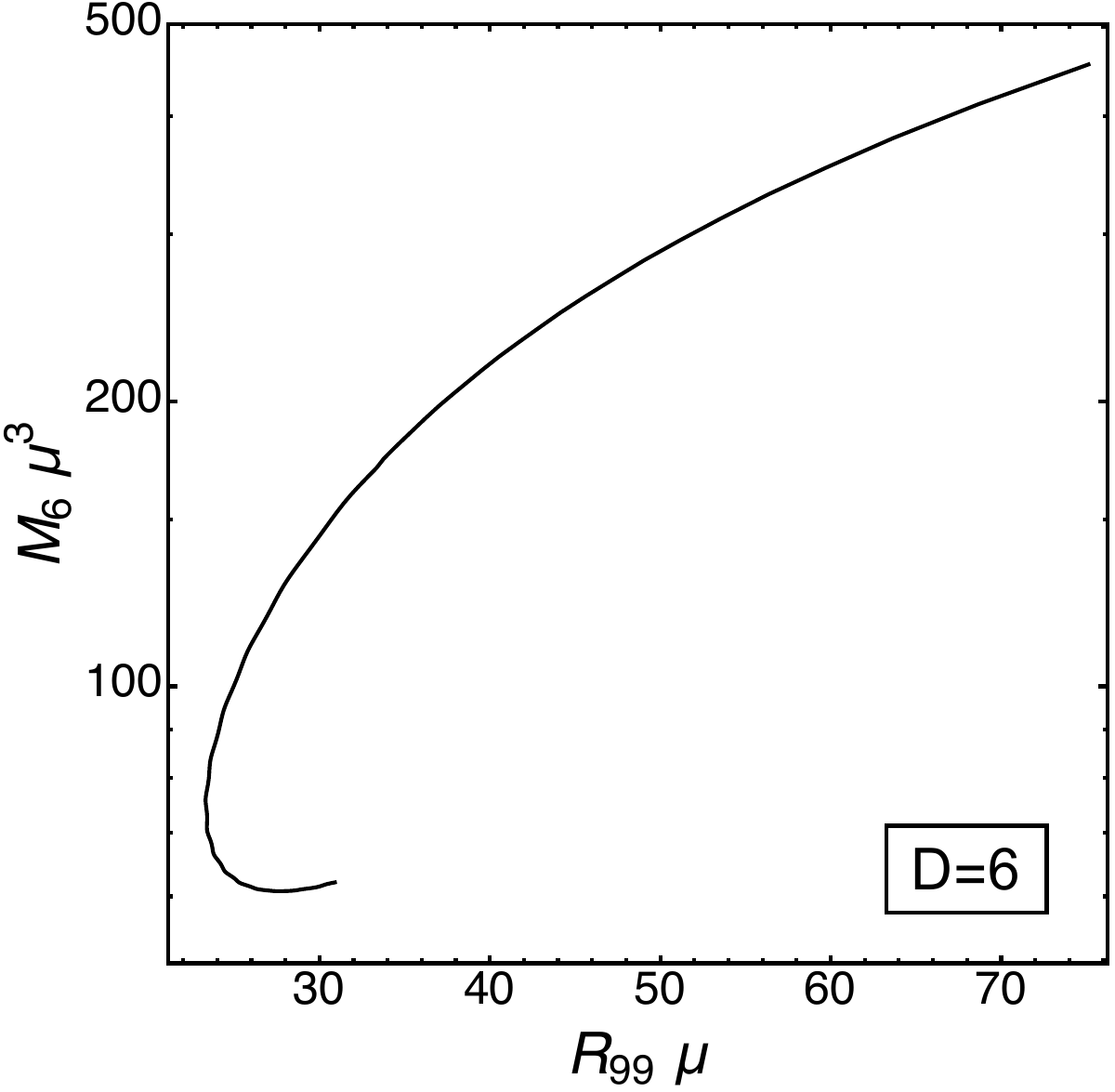}
\caption{Total mass as a function of the effective radius $R_{99}$ --- defined as the radius within which 99\% of $M_D$ is contained --- for mini-boson-star equilibrium solutions in $D=\{4,5,6\}$.}
\label{fig:massradius}
\end{figure}

A useful and important information is how the total mass of boson stars changes with the frequency of the ground state, illustrated in the typical spiralling diagrams in \cref{fig:massfrequency} for $D=\{4,5,6\}$.
A complementary piece of information is shown in \cref{fig:massradius}, where we plot the total mass as a function of the effective radius of the boson star.
In four dimensions, the total mass has a maximum (the Kaup limit, $M_{4,\max}\approx0.633/\mu$) for a finite value of $\phi_c$, and vanishes for $\phi_c\to0$, or equivantly $\omega\to\mu$.
Instead, in higher dimensions, it seems that the maximum mass corresponds to the limiting case $\phi_c\to0$, implying a mass gap that disconnects higher dimensional solutions from the Minkowski spacetime.
We also notice that as the number of dimensions increases, the range of frequencies which allows for solutions narrows towards $\omega=\mu$.
Modulo different conventions and normalisations, our results are in agreement with other published results~\cite{Hartmann:2010pm,Brihaye:2015jja,Brihaye:2016ibz,Blazquez-Salcedo:2019qrz,Astefanesei:2003qy}.
Background solutions can be found for higher dimensions, but they require high numerical precision and fine tuning of the shooting parameters.

We have verified that the maximum mass $M_{D,\max}$ and maximum boson number $N_{D,\max}$ diverge as $\e^{\alpha D}$ with $\alpha$ some positive number, as $\phi_c\to0$. The binding energy is always positive.

\section{Stability analysis}

The stability properties are studied via an analysis of how the system behaves under small fluctuations. But before proceeding with the full analysis, a few observations are in order.

\begin{figure}[ht]
\centering
\includegraphics[height=0.27\textwidth]{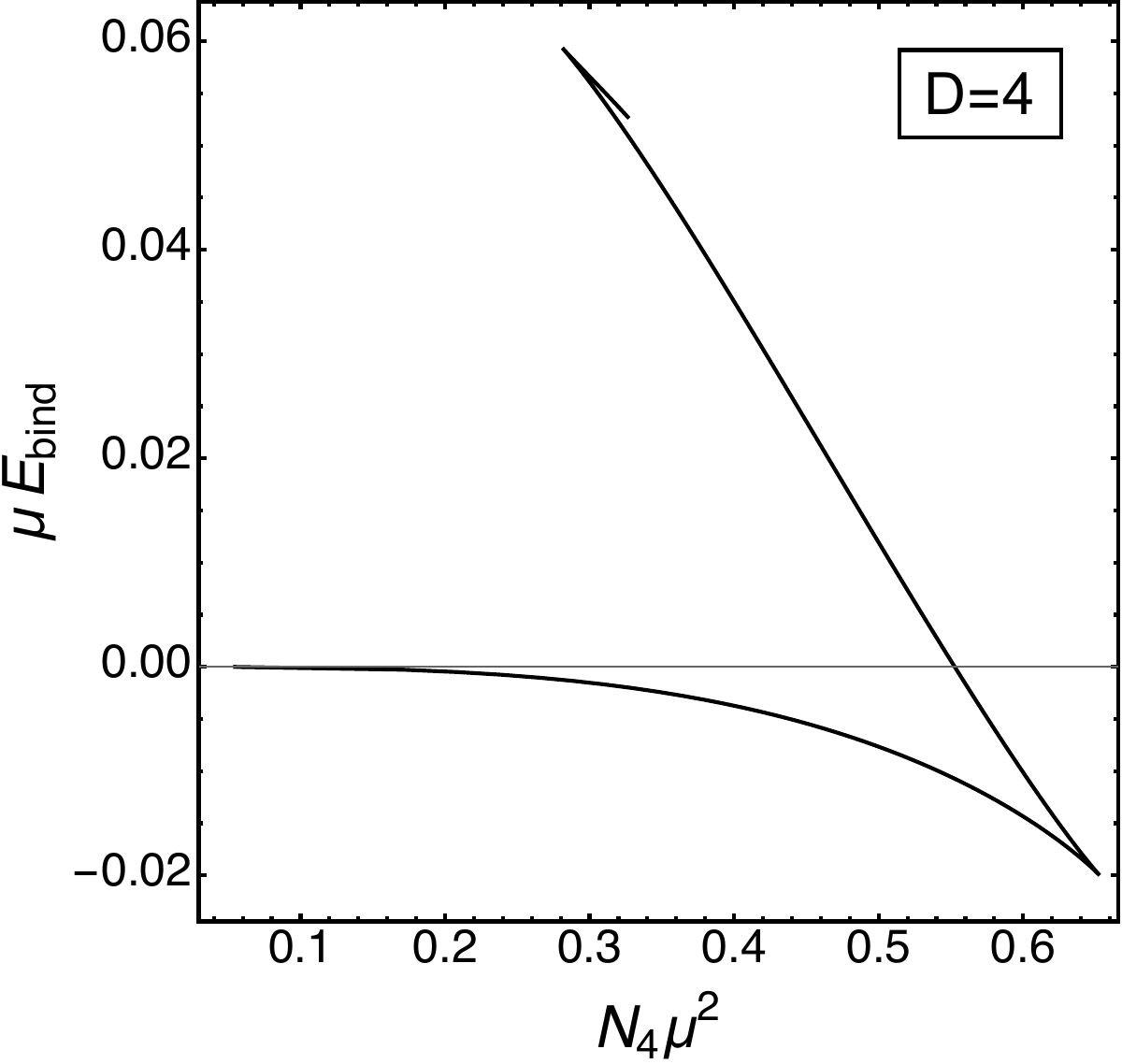}\quad
\includegraphics[height=0.27\textwidth]{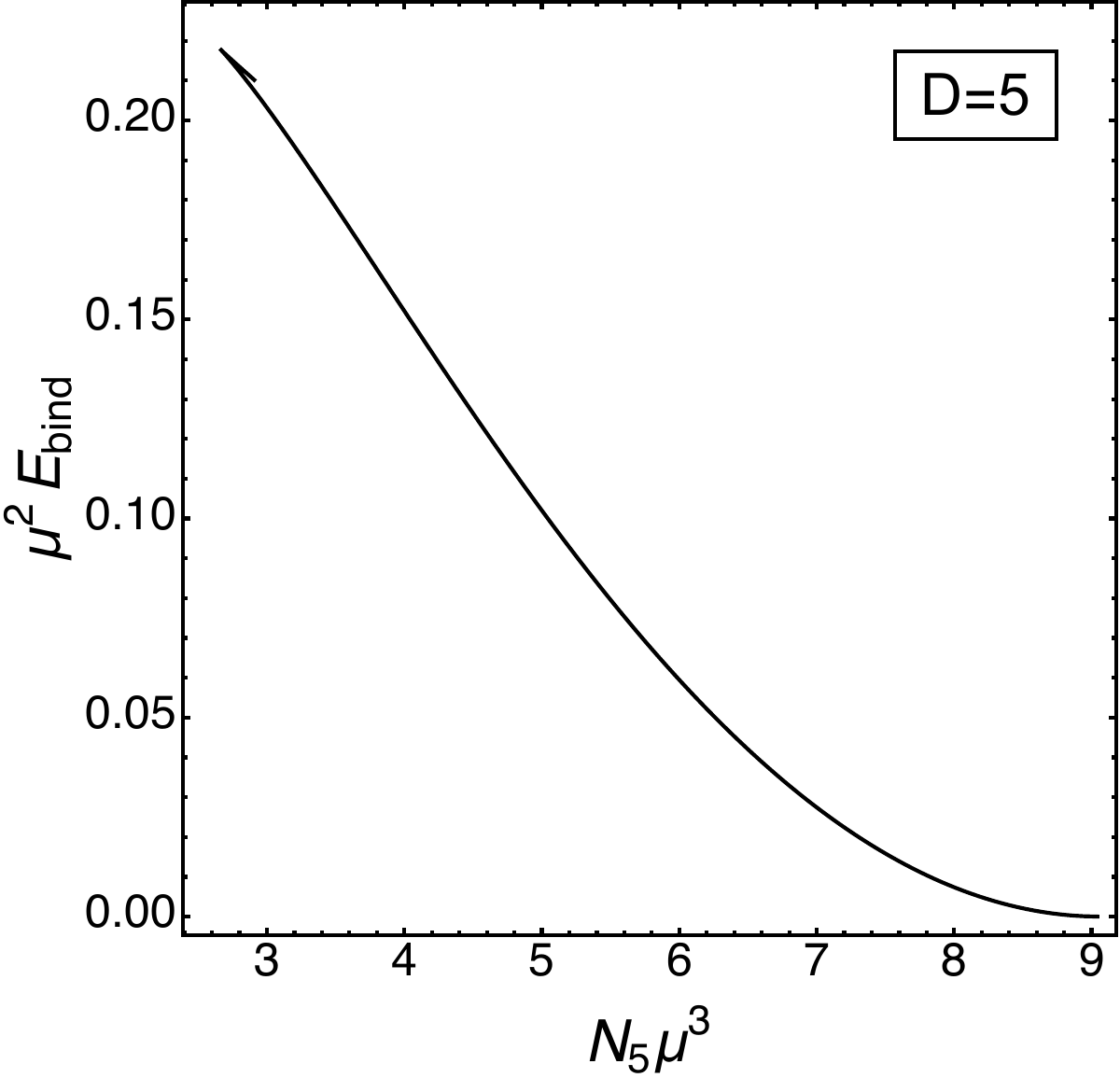}\quad
\includegraphics[height=0.27\textwidth]{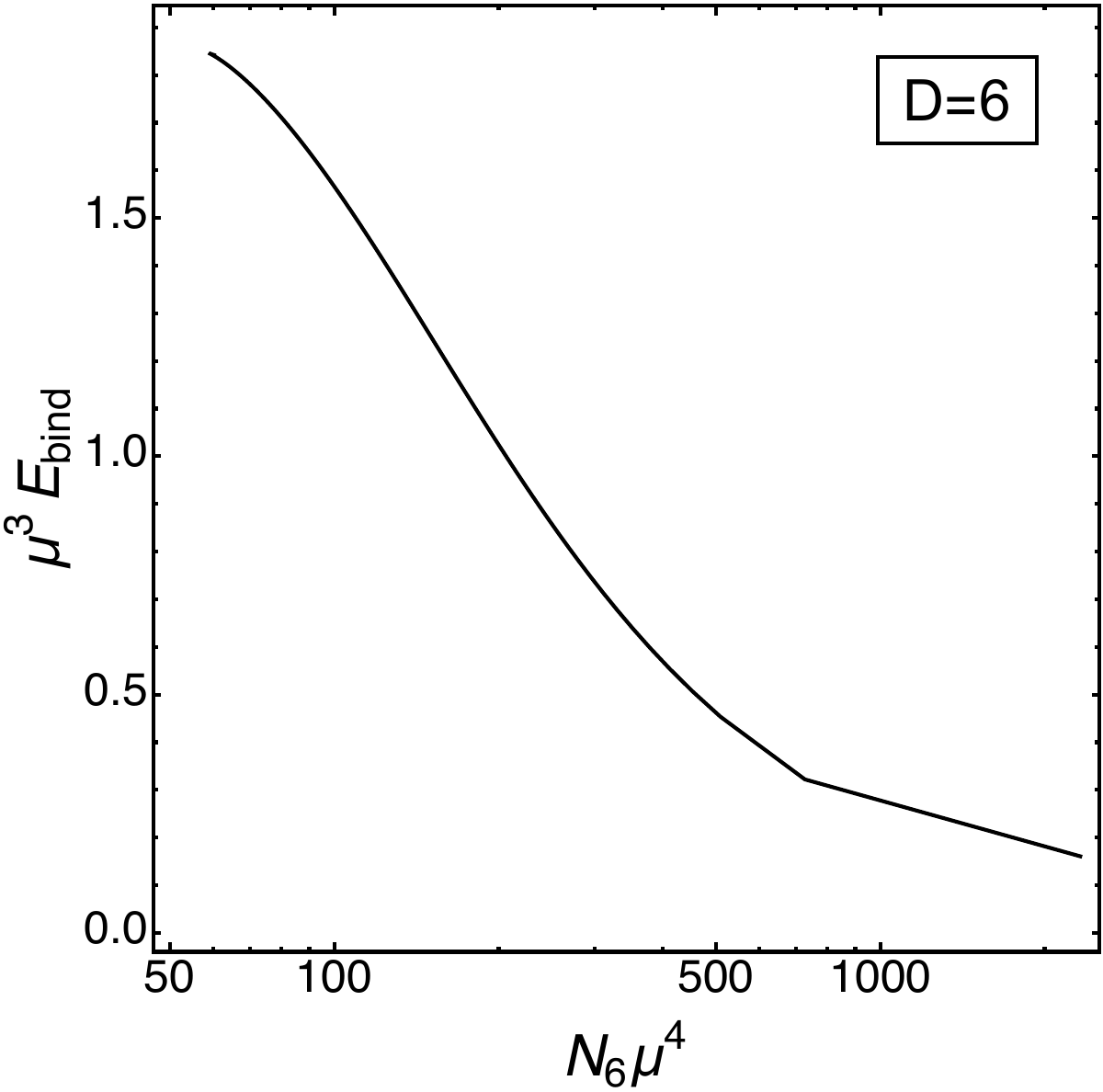}
\caption{Bifurcation diagrams of the binding energy versus the total boson number for mini-boson-star equilibrium solutions in $D=\{4,5,6\}$.}
\label{fig:bindenergy}
\end{figure}

The existence of a maximum mass is usually an indication of a point of marginal stability~\cite{Gleiser:1988rq}.
Negative binding energy is a necessary but not sufficient condition to have stable configurations.
It follows that the stability of the solutions can also be studied employing a binding-energy stability criterion~\cite{Kusmartsev:1990cr}, by looking at the behavior of the binding energy as a function of the boson number, shown in \cref{fig:bindenergy} for equilibrium solutions in $D=\{4,5,6\}$.
Therefore, we expect a stable branch in $D=4$, as confirmed e.g., in Refs.~\cite{Hawley:2000dt,Kain:2021rmk,Santos:2024vdm}, and instability for $D>4$.

In four dimensions, the cusp occurs at the location of the maximum mass and marks the onset of instability~\cite{Kusmartsev:1990cr}. The line branching off at that point, with steeper slope, corresponds to boson stars with a larger value of the scalar at the origin, and eventually reaches a positive binding energy.
In higher dimensions the binding energy is always positive, providing a strong hint of instability.

To study linear stability, we follow Refs.~\cite{Gleiser:1988ih,Hawley:2000dt} and we expand about the equilibrium configuration: $\lambda(t,r)=\lambda(r)+\delta\lambda(t,r)$, $\nu(t,r)=\nu(r)+\delta\nu(t,r)$, and $\Phi(t,r)=\frac{1}{\sqrt{8\pi}}\,\e^{-\iu\omega t}\left[\phibg(r)+\delta\phi(t,r)\right]$.
These perturbations preserve spherical symmetry and will give rise to radial oscillations.
The Einstein--Klein--Gordon equations for the perturbed quantities can be manipulated such that we get rid of $\delta\nu$, and we are left with two equations in two unknowns,
\be
\delta\lambda'' &= \e^{\lambda-\nu}\,\ddot{\delta\lambda} + \left[\frac{2 (D-3)}{r^2}-\frac{2 (D-3) \nu'+\lambda'}{r}-\frac{\left(\nu'-\lambda'\right)^2}{2} +\lambda'' + 2 \phibg'^2\right]\!\delta\lambda \0\\
&\phantom{=} + \left[\frac{3\left(\lambda'-\nu'\right)}{2} -\frac{D-4}{r}\right]\!\delta\lambda'
+\frac{2\mu^2 \e^{\lambda} r \left[\phibg \left(\lambda'+2 \nu'\right) + 2 \phibg'\right]}{D-2}\,\delta\phi
+ 4\left[\frac{\e^{\lambda} \mu^2 r \phibg}{D-2} - \phibg'\right]\!\delta\phi',\label{deltalambda}\\
\delta\phi'' &= \e^{\lambda -\nu}\,\ddot{\delta\phi} + \e^{\lambda} \left[\frac{2 \mu^2 r \phibg \phibg'}{D-2}+3 \mu^2+\e^{-\nu} \omega^2\right]\!\delta\phi + \left[\frac{\lambda'-\nu'}{2}-\frac{D-2}{r}+\frac{2 \phibg'}{\phibg}\right]\!\delta\phi'\0\\
&\phantom{=} - \e^{\lambda} \left[\frac{(D-3) (D-2)}{r^2 \phibg}-\frac{\mu^2 r \phibg^2 \phibg'}{D-2}+\frac{(D-3) \phibg'}{r}-2 \mu^2 \phibg\right]\!\delta\lambda-\frac{D-2}{r \phibg}\,\delta\lambda'\,.\label{deltaphi}
\ee

Assuming harmonic time dependence, i.e., $\delta\phi(t,r)=\delta\phi(r)\,\e^{\iu\sigma{}t}$ and $\delta\lambda(t,r)=\delta\lambda(r)\,\e^{\iu\sigma{}t}$, the system defined by \cref{deltaphi,deltalambda} along with the condition that the boson number must be conserved, 
%
%
establishes a characteristic value problem for $\sigma^2$.
It can be shown that the system is self-adjoint, hence the values of $\sigma^2$ must be real.
To determine stability or instability we just need to determine whether $\sigma^2$ is positive or negative.
Negative values of $\sigma^2$ correspond to perturbations that will grow which means that the boson star is unstable against radial oscillations.

The first terms of the regular boundary conditions at $r=0$ are
\begin{subequations}\be
\delta\lambda &= 0\,,\quad \delta\lambda'=0\,,\quad \delta\lambda''=\delta\lambda_c\,,\\
\delta\phi &= \delta\phi_c\,,\quad \delta\phi'=0\,,\quad
\delta\phi'' = \frac{3 \mu^2 - \e^{-\nu_c} \left(\sigma^2 - \omega^2\right)}{D-1}\,\delta\phi_c - \frac{D-2}{2\phi_c}\,\delta\lambda_c\,.
\ee\end{subequations}
At infinity the perturbations vanish, i.e., $\delta\lambda=\delta\phi=0$.

To solve \cref{deltaphi,deltalambda} for a given $\phi_c$ we begin by solving the respective background solution and then we shoot for $\sigma^2$ and $\delta\lambda_c$.
For any given background configuration, there might be a discrete number of values for $\sigma^2$; we only look for the fundamental mode as if it is stable, all radial modes will be, while if it is unstable, it corresponds to the fastest growing instability.
Notice that, since the system is linear, the value of $\delta\phi_c$ is arbitrary.

\begin{figure}[!hbt]
\centering
\includegraphics[height=0.27\textwidth]{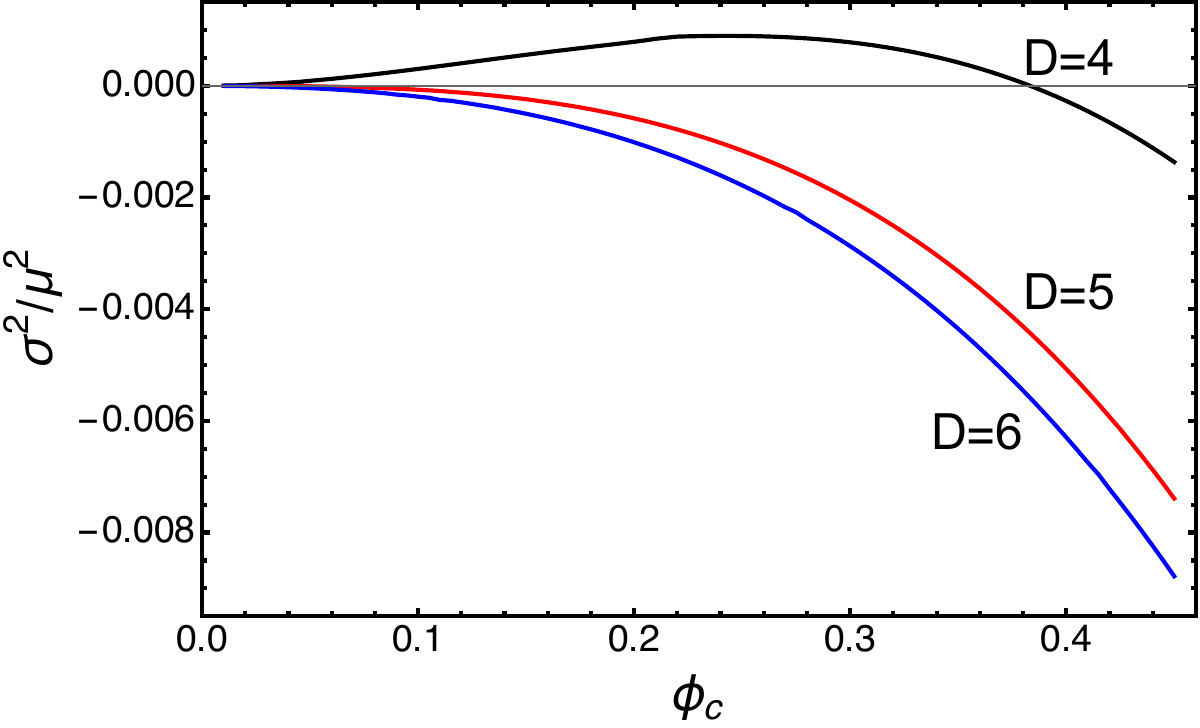}
\caption{Distribution for the squared frequency $\sigma^2$ with respect to the central value of the scalar field $\phi_c$.
In four dimensions $\sigma^2$ crosses zero for the value of $\phi_c$ that corresponds to the maximum mass configuration, while in five and six dimensions $\sigma^2$ is always negative.}
\label{fig:instability}
\end{figure}

In \cref{fig:instability} we plot the dependence of the squared frequency $\sigma^2$ on the scalar field central value $\phi_c$ for mini boson stars in $D=\{4,5,6\}$.
The values of $\phi_c$ depend on units and conventions, and plotting against $\phi_c$ might not seem the best choice.
Despite this, quantities such the total mass or the frequency vary in very different ranges subject to the dimensionality of the spacetime, making the comparison less readable.
We observe that $\sigma^2$ is always negative in the higher-dimensional cases, contrary to the four-dimensional case~\cite{Hawley:2000dt}, implying that mini-boson-star solutions in higher spacetime dimensions are always unstable against radial perturbations.
Moreover, the absolute value of $\sigma^2$ is larger for $D=6$ than for $D=5$, so we expect that boson stars in higher dimensions are hierarchically more unstable.

\section{Conclusions}

Dimensionality can affect the physical consequences of a given theory.
In this note we have studied the existence and stability of mini boson stars in $D$ non-compact dimensions, proving through a dynamical analysis of the perturbation equations that they are always unstable against radial oscillations for $D>4$.

This result strongly confirms weaker statements in terms of the kinetic and gravitational energy balance, and binding-energy arguments.

In higher dimensions, different self-interacting terms, different couplings with gravity, and even different spacetime asymptotics, could sustain gravitational pressure, give configurations with negative binding energy, and perhaps allow for a stable branch. This however depends on the fine details of the theory and requires an extensive analysis which is left for future work.

\bigskip
\bmhead*{Acknowledgments}
\addcontentsline{toc}{section}{Acknowledgments}
I thank V\'itor Cardoso for several discussions long time ago.
I acknowledge funding support from the MUR PRIN (Grant 2020-KR4KN2 ``String Theory as a bridge between Gauge Theories and Quantum Gravity''), FARE (GW-NEXT, CUP:\ B84I20000100001) programmes, and from EU Horizon 2020 Research and Innovation Programme under the Marie Sk\l{}odowska-Curie Grant Agreement no.\ 101007855.

\medskip
\bibliographystyle{sn-nature}
\bibliography{refs}
\addcontentsline{toc}{section}{References}

\end{document}